\documentclass[twocolumn]{aastex61}

\usepackage{amsmath}

\newcommand{\beq}{\begin{equation}}
\newcommand{\eeq}{\end{equation}}

\newcommand\reye{\mathrm{Re}}
\newcommand\reym{\mathrm{Rm}}
\newcommand{\Pm}{\mathrm{Pm}}
\newcommand{\Co}{\mathrm{Co}}

\submitjournal{ApJ}

\shorttitle{The weakly nonlinear magnetorotational instability in a local geometry}
\shortauthors{Clark \& Oishi}

\begin{document}

\title{The Weakly Nonlinear Magnetorotational Instability in a local geometry}

\correspondingauthor{S. E. Clark}
\email{seclark@astro.columbia.edu}

\author[0000-0002-7633-3376]{S. E. Clark}
\affil{Department of Astronomy, Columbia University, New York, NY}

\author[0000-0001-8531-6570]{Jeffrey S. Oishi}
\affiliation{Department of Physics and Astronomy, Bates College, Lewiston, ME}
\affiliation{Department of Astrophysics, American Museum of Natural History, New York, NY}
\nocollaboration

\begin{abstract}

The magnetorotational instability (MRI) is a fundamental process of accretion disk physics, but its saturation mechanism remains poorly understood despite considerable theoretical and computational effort. We present a multiple scales analysis of the non-ideal MRI in the weakly nonlinear regime -- that is, when the most unstable MRI mode has a growth rate asymptotically approaching zero from above. Here, we develop our theory in a local, Cartesian channel. Our results confirm the finding by Umurhan et al. (2007) that the perturbation amplitude follows a Ginzburg-Landau equation. We further find that the Ginzburg-Landau equation will arise for the local MRI system with shear-periodic boundary conditions when the effects of ambipolar diffusion are considered. A detailed force balance for the saturated azimuthal velocity and vertical magnetic field demonstrates that even when diffusive effects are important, the bulk flow saturates via the combined processes of reducing the background shear and rearranging and strengthening the background vertical magnetic field. We directly simulate the Ginzburg-Landau amplitude evolution for our system and demonstrate the pattern formation our model predicts on long length and time scales. We compare the weakly nonlinear theory results to a direct numerical simulation of the MRI in a thin-gap Taylor Couette flow.

\end{abstract}

\section{Introduction} \label{sec:intro}
For matter to accrete from a disk onto a central object, angular momentum must be transported radially outward in the disk. The transport mechanism is likely turbulent, as molecular viscosity alone cannot account for the needed angular momentum transfer, and likely magnetic, as this turbulence is excited even in hydrodynamically stable disks \citep{Shakura:1973wg}. Discovered by \citet{Chandrasekhar:1960wh} and \citet{Velikhov:1959} in a global geometry, the magnetorotational instability (MRI) was subsequently rediscovered and applied to accretion disks by \citet{Balbus:1991vs}. Since then, the MRI remains the leading explanation for rapid angular momentum transport in astrophysical disks. The instability in its simplest geometry arises when a differentially rotating disk is threaded by a vertical magnetic field. The presence of the magnetic field linearly destabilizes the disk gas, driving turbulence and angular momentum transport \citep[e.g.][]{2011ApJ...738...84H,2014MNRAS.438.2513P,2013MNRAS.435.2281P}. The MRI likely plays a role in a diverse host of astrophysical systems, including protoplanetary disks \citep[e.g.][]{2015ApJ...798...84B} and black hole accretion disks \citep[e.g.][]{2013ApJ...769..156S}, as well as stellar interiors \citep[e.g.][]{2015ApJ...799...85W}. Despite its importance, many aspects of the MRI remain poorly understood. In particular, the nonlinear saturation mechanism for the MRI is an open question, and a formidable challenge. MRI saturation has been tackled almost exclusively with simulation, with a few notable exceptions detailed below. In this work we analytically investigate the weakly nonlinear saturation of the MRI.

Weakly nonlinear analysis is a perturbative method used to examine the asymptotic behavior of a system near threshold -- that is, when the system is just barely unstable to its most unstable mode. The analytical technique follows the multiscale evolution of fluid variables in a perturbation expansion, allowing the controlled interaction of modes between orders in a perturbation series \citep{1978amms.book.....B}. Weakly nonlinear analysis can be a powerful technique for analytically examining systems which in their full generality exhibit such complicated nonlinear behavior that their study is relegated primarily to the simulation domain. The MRI is one such phenomenon: while there is a rich literature analytically examining the linear MRI, analytical treatments of the \textit{nonlinear} system are relatively few. The weakly nonlinear treatment of the MRI was pioneered by \citet{Knobloch:2005ba} and \citet[][hereafter URM07; see also \citeyear{Umurhan:2007dz}]{Umurhan:2007hs}. The latter authors undertook the first weakly nonlinear analysis of the MRI in a thin-gap Taylor-Couette (TC) flow with strong dissipation (as is appropriate to liquid metal experiments), and found that the marginal MRI system approaches saturation in a manner analogous to that of Rayleigh-B\'enard convection. Weakly nonlinear analysis was instrumental in our understanding of Rayleigh-B\'enard convection saturation \citep{Newell:1969}, and the similarities between convection and the local MRI are the result of important shared symmetries between the systems. The success of URM07 in modeling the MRI system near threshold merits further consideration, but we are unaware of any other attempts to expand upon their theoretical framework. In this work we rederive the theory of URM07, and expand upon their findings. Our focus here is on fully characterizing the local MRI system, both for independent theoretical interest and to have a robust comparison point for extensions of this theory into more complicated geometries. In a companion paper we derive for the first time the weakly nonlinear theory of the standard and helical MRI in a global, cylindrical TC flow \citep[][hereafter Paper II]{Clark:2016}. The thin- and wide-gap treatments complement one another theoretically, and both are important regimes for comparison with simulation. 

This work examines TC flow in the thin-gap regime, an idealization in which the radial extent of the channel is very small compared to its distance from the center of rotation, i.e. $(r_2 - r_1) \ll \frac{1}{2} (r_1 + r_2)$ where $r_1$ and $r_2$ are the radii of the inner and outer flow boundaries, respectively. The thin-gap approximation eliminates curvature terms, so the domain geometry is Cartesian rather than cylindrical. The excluded curvature terms have an explicit dependence on $r$, so they make the problem more challenging both analytically and numerically. In particular, in the wide-gap geometry (i.e. true Taylor-Couette flow) the base angular velocity is a function of $r$, where in the thin-gap approximation the shear flow reduces to a linear profile. The equations of motion in thin-gap TC flow are thus identical to the MRI in a local shearing box, which differs from our fiducial setup only in the application of periodic boundary conditions.

We note several other important analytical studies of MRI saturation. \citeauthor{Knobloch:2005ba} (\citeyear{Knobloch:2005ba}) analyze the MRI in the strongly nonlinear regime, by following the already-developed MRI modes into asymptotic saturation. They consider a thin-gap regime as well, and so their theory may be considered the strongly nonlinear analogue to the one developed here. \citet{Vasil:2015} examines the weakly nonlinear MRI in a thin-gap regime in a minimal model, finding deep mathematical similarities between the MRI system and the elastodynamic instability of a buckling beam. We discuss these results and their relation to ours in Section~\ref{sec:discussion}. Several authors have investigated the behavior of the MRI when the boundary conditions are shear periodic, and so the MRI has no mechanism by which to modify the background shear flow profile. In this approximation linear MRI growth is dominated by channel modes, a type of MRI mode that, for periodic boundary conditions, are exact solutions of both the linear and nonlinear MRI equations \citep{Goodman:1994ul}. In this regime the MRI saturates via parasitic instabilities, which feed off and destroy the primary MRI modes. Analytical investigation of this case reveals that MRI saturation can be caused by parasitic Kelvin-Helmholtz and tearing mode instabilities, depending on parameter regime \citep{Pessah:2010ic}. The theory of MRI channel mode parasites is robust \citep[see also][]{Pessah:2009gm,Latter:2010iz,Rembiasz:2016}, but their importance may be overestimated by the local approximation \citep{Latter:2015}, and not germane to global analyses like the one presented here. \citet{Latter:2015} gives a detailed analysis of the relation between local and global linear MRI modes. In this work we describe the applicability of our weakly nonlinear theory to shear-periodic boxes. We find that under certain conditions the weakly nonlinear mode interaction described here may provide an alternative MRI saturation mechanism in the shearing box that does not rely on parasitic instabilities.

We begin with an overview of our basic model equations for the local MRI in Section~\ref{sec:equations} and then describe our weakly nonlinear analysis and give results for the thin-gap TC flow in Section~\ref{sec:wnl}. In Section \ref{sec:SBAD} we detail the conditions under which our theory applies to the case where the boundary conditions are shear periodic, namely the consideration of ambipolar diffusion. We compare our results to a direct numerical simulation in Section \ref{sec:simulations}. We then place our results in the context of previous results from both analytic and computational studies in Section~\ref{sec:discussion} and draw conclusions in Section~\ref{sec:conclusion}.

\begin{figure}[h!]
\centering
\includegraphics[trim=5cm 0cm 0cm 0cm, scale=.35]{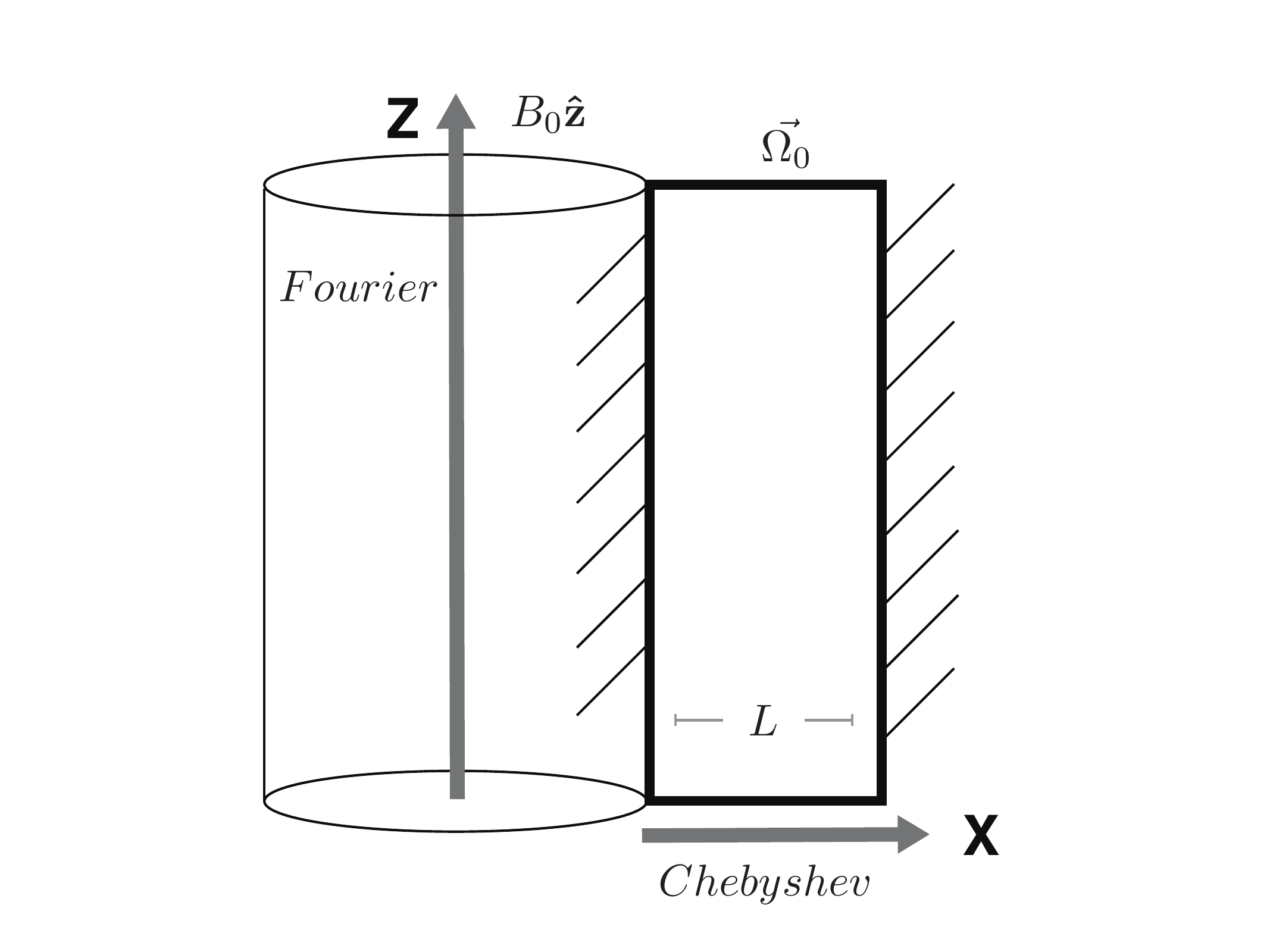}
\caption{Schematic diagram of our set-up, an axisymmetric thin-gap Taylor-Couette flow. We investigate a 2D slice of the X-Z (radial-vertical) plane. Our domain is represented by the bolded black box, of width L. The radial dimension is solved with a basis of Chebyshev polynomials, and the vertical dimension is solved on a Fourier basis. }\label{fig:setup}
\end{figure}

\section{Equations}\label{sec:equations}

The evolution of a conducting fluid is governed by the momentum and induction equations,

\begin{eqnarray}\label{eq:momentum}
\begin{split}
& \partial_t \mathbf{u} \, + \, \mathbf{u} \cdot \nabla \mathbf{u} \, = \, -\frac{1}{\rho}\nabla P \, - \, \nabla\Phi \, + \, \frac{1}{\rho} \left(\mathbf{J}\times\mathbf{B}\right) \, \\
& + \, \nu\nabla^2 \mathbf{u} \, - \, 2\mathbf{\Omega} \times \mathbf{u} \, - \, \mathbf{\Omega} \times \left(\mathbf{\Omega} \times \mathbf{r} \right), \\
\end{split}
\end{eqnarray}

\beq\label{eq:induction}
\partial_t \mathbf{B} = \nabla \times \left(\mathbf{u} \times \mathbf{B}\right) + \eta\nabla^2\mathbf{B},
\eeq

where $P$ is the gas pressure, $\nu$ is the kinematic viscosity, $\eta$ is the microscopic diffusivity, $\nabla\Phi$ is the gravitational force per unit mass, and the current density is $\mathbf{J} = \nabla\times\mathbf{B}$.
Equations \ref{eq:momentum} and \ref{eq:induction} are subject to the incompressibility and magnetic solenoid constraints,

\beq\label{eq:incompressibility_constraint}
\nabla \cdot \mathbf{u} = 0
\eeq

\beq\label{eq:solenoid_constraint}
\nabla \cdot \mathbf{B} = 0.
\eeq

We axisymmetrically perturb all three vector components of each of the fluid quantities. We nondimensionalize the equations, with lengths nondimensionalized by $L$, time by $\Omega_0$, velocities by $\Omega_0 L$, magnetic fields by $B_0$, and pressure by $\Omega_0^2 L^2 \rho_0$, where $L$ is the channel width, $\Omega_0$ is the rotation rate at the center of the channel, and $\rho_0$ is the constant pressure in the base state (see Figure \ref{fig:setup}). We define the Reynolds number, $\reye \equiv {\Omega_0 L^2}/{\nu}$, magnetic Reynolds number, $\reym \equiv {\Omega_0 L^2}/{\eta}$, and Cowling number, $\Co \equiv {2 v_A^2}/{\Omega_0^2r_0^2}$, where the Alfv\'en speed $v_A$ is $v_A^2 = {B_0^2}/{\rho_0}$. The fluid symbols $\mathbf{u}$, $\mathbf{B}$, etc. will henceforth be used to refer to the nondimensional, perturbed quantities.

We define the streamfunction $\Psi$ and flux function $A$, where $A$ is the familiar two-dimensional vector potential. $\Psi$ and $A$ are scalar fields. The curl of $\Psi$ and the curl of $A$ are defined as the velocity and magnetic field perturbation, respectively, and so $\Psi$ and $A$ automatically satisfy our constraints (Equations \ref{eq:incompressibility_constraint} and \ref{eq:solenoid_constraint}).

$\Psi$ and $A$ are thus related to the velocity and magnetic field perturbations, respectively, as

\beq
\mathbf{u} \, = \, \left[\begin{matrix}
\partial_z \Psi \\
u_{y} \\
-\partial_x \Psi \\
\end{matrix}\right],\eeq

\beq
\mathbf{B} \, = \, \left[\begin{matrix}
\partial_zA \\
B_{y} \\
-\partial_xA \\
\end{matrix}\right].\eeq

\onecolumngrid
Our final equation set is

\beq
\label{eqset1}
\partial_t \nabla^2 \Psi \, - \, 2 \partial_z u_{y} \, - \, \Co B_0 \partial_z \nabla^2 A \, - \, \frac{1}{\reye}\nabla^4 \Psi \, = \, \Co J\left(A, \nabla^2 A \right) \, - \, J\left(\Psi, \nabla^2 \Psi\right) \, 
\eeq

\beq
\label{eqset2}
\partial_t u_{y} \, + \, \left(2 - q\right) \Omega_0 \partial_z \Psi \, - \, \Co B_0\partial_z B_{y} \, - \, \frac{1}{\reye} \nabla^2 u_{y} \, = \, \Co J\left(A, B_{y}\right) \, - \, J\left(\Psi, u_{y}\right) 
\eeq

\beq
\label{eqset3}
\partial_t A \, - \, B_0 \partial_z \Psi \, - \, \frac{1}{\reym} \nabla^2 A \, = \, J\left(A, \Psi\right) 
\eeq

\beq
\label{eqset4}
\partial_t B_{y} \, + \, q \Omega_0 \partial_z A \, - \, B_0 \partial_z u_{y} \, - \, \frac{1}{\reym} \nabla^2 B_{y} \, = \, J\left(A, u_{y}\right) \, - \, J\left(\Psi, B_{y}\right)  ,
\eeq
\twocolumngrid

where $J$ is the Jacobian operator, 
\beq
J\left(f, g\right) \equiv \partial_z f\partial_x g - \partial_x f \partial_z g,
\eeq  
and $q \equiv - d \ln \Omega/ \ln R = 3/2$ is the dimensionless shear parameter defining a rotation profile $\Omega(r) = \Omega_0 (r/r_0)^{-q}$, such that the background velocity profile is $u_0 = -q \Omega_0 x$.

\begin{figure}[h!]
  \centering
  \includegraphics[width=\columnwidth]{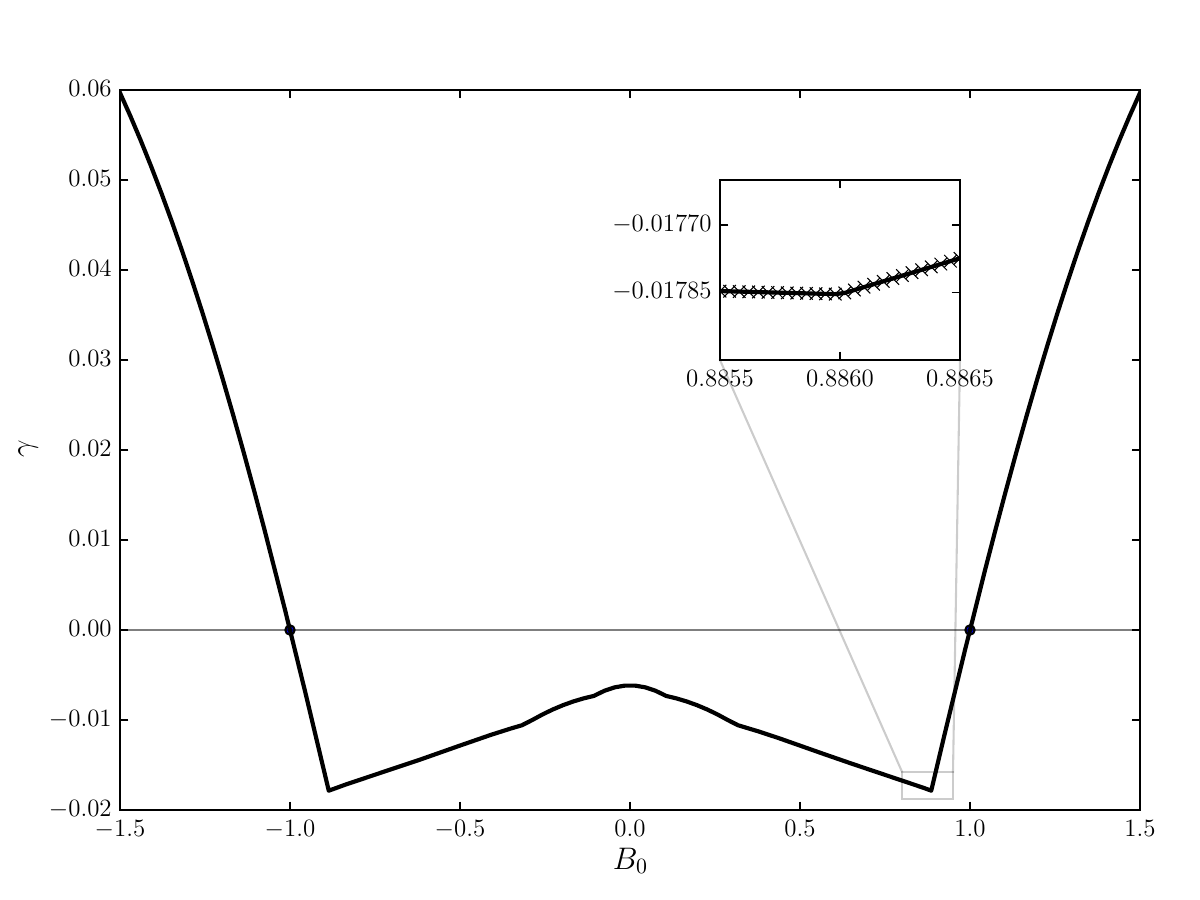}
  \caption{Growth rate $\gamma$ as a function of background magnetic field strength $B_0$ at $\reym = \reym_c$, $k_z = k_c$, $\Pm = 10^{-3}$. Around the critical value $B_0 = 1.$, \emph{strengthening} $B_0$ tunes the system into instability, while \emph{decreasing} it leads to stability. The inset highlights the fact that $\gamma$ is determined by the maximum real eigenvalue of the system, which switches from one mode family to another as discussed in the text.}
  \label{fig:growth_rate_vs_B0}
\end{figure}

The weakly nonlinear regime is where the MRI system is nonlinearly unstable to only the most unstable mode of the linear solution. We find the marginal state, where the most unstable linear MRI mode neither grows nor decays, for a set of dimensionless parameters, and then destabilize the system. We examine the system for fiducial parameters comparable to URM07, namely $\Pm = 1.0 \times 10^{-3}$, $\Co = 0.08$, $q = 1.5$. The system is marginal for a critical wavenumber $k_c = 0.75$ and a critical magnetic Reynolds number $\reym_c = 4.9$.

\begin{figure*}
\centering
\includegraphics[width=\textwidth]{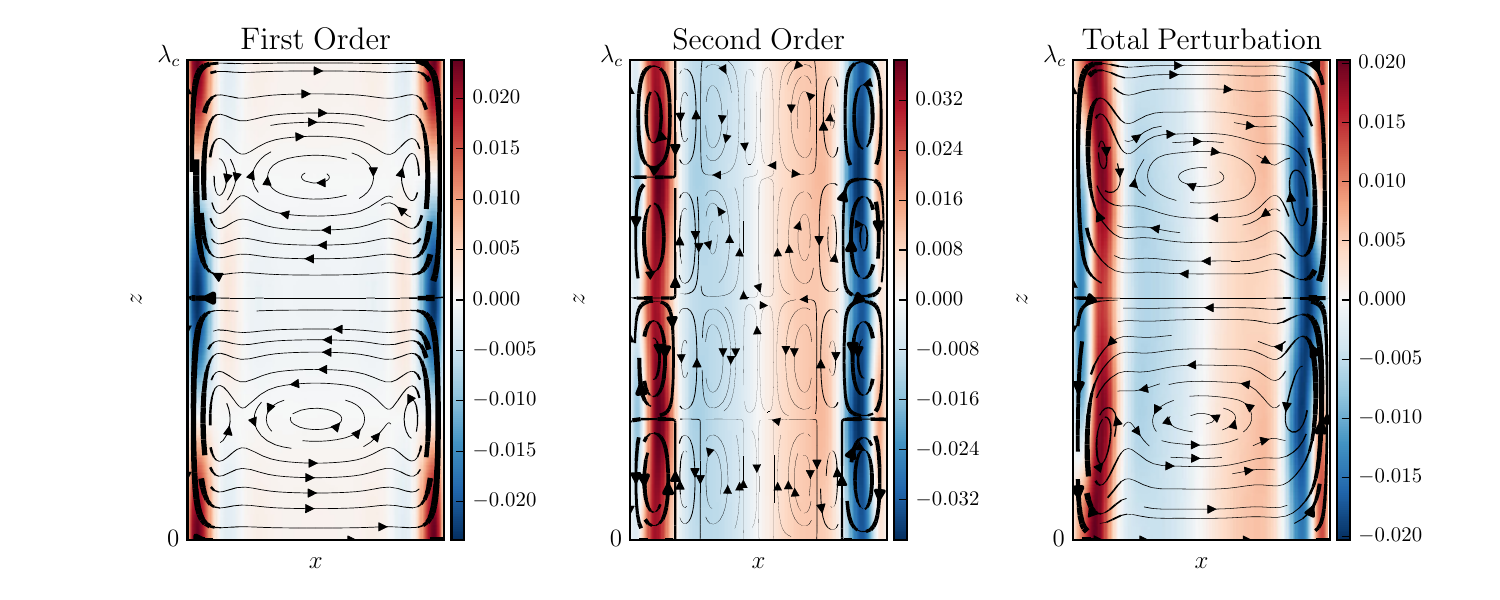}
\caption{First order (left), second order (center), and total (right) velocity perturbations. Streamlines represent velocity in the vertical-radial plane, where thicker streamlines correspond to faster speeds. Colorbar represents azimuthal velocity. We use a constant amplitude $\alpha = \alpha_{saturation}$ and a small parameter $\epsilon = 0.5$.}\label{fig:allorders_velocity}
\vspace*{\floatsep}
\centering
\includegraphics[width=\textwidth]{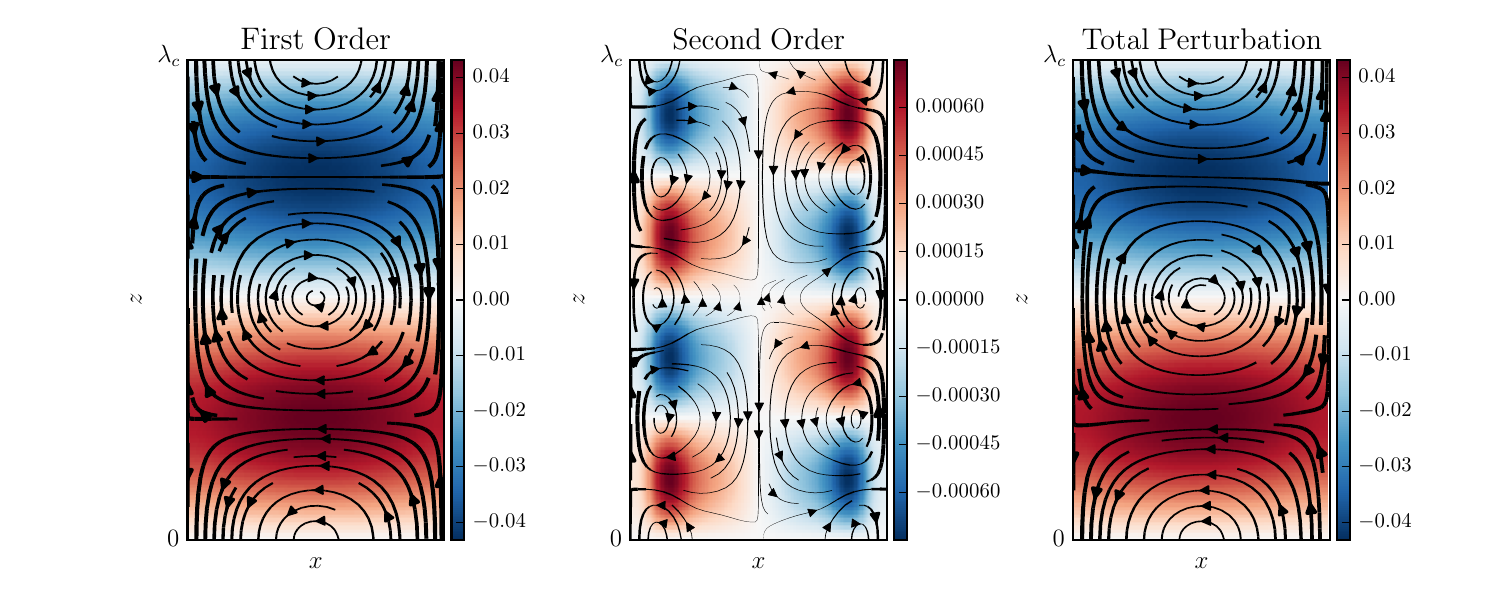}
\caption{As in Figure \ref{fig:allorders_velocity} but for the magnetic field. Streamlines represent magnetic field structure in the vertical-radial plane, where thicker streamlines correspond to higher magnetic field strengths. Colorbar represents azimuthal magnetic field strength.}\label{fig:allorders_Bfield}
\end{figure*}

Because we nondimensionalize $B$ by the magnitude of the background field strength, $B_0 \equiv 1$ in Equations \ref{eqset1} - \ref{eqset4}. To excite the weakly nonlinear MRI, we tune the background magnetic field away from stability. We do so by substituting $B = B_0\left(1 + \epsilon^2\right)$. The degree of departure from the marginal state is measured by the small parameter $\epsilon$. An $\mathcal{O}\left(\epsilon^2\right)$ strengthening of the background magnetic field destabilizes a finite band of wave modes with a width of $\mathcal{O}\left(\epsilon\right)$, which interact nonlinearly. We note that this definition of $\epsilon$ is opposite in sign to nearly all previous works \citep[e.g.][]{Umurhan:2007dz,Umurhan:2007hs}. Because in the ideal limit, the MRI can be tuned into instability by setting $B_0$ to its critical value and then \emph{decreasing} its value, it is natural to consider $\epsilon^2$ as a weakening of the background field \citep[as is done correctly in][for example]{Vasil:2015}. However, as we show in figure~\ref{fig:growth_rate_vs_B0}, for the \emph{dissipative} case with $\eta, \nu \ne 0$, when all other parameters are critical, decreasing $B_0$ leads to stability, while increasing it pushes the system into instability. Figure~\ref{fig:growth_rate_vs_B0} is symmetric about $B_0 = 0$, as it must be, since the MRI is insensitive to the sign of the background field. There are several places at which the derivative of $\gamma$ appears discontinuous; this is not physical but rather reflects the fact that we define $\gamma$ as the growth rate of the most unstable mode. That is, it is the maximum real part of the eigenvalues of the linearized system (e.g. equation~\ref{eq:unperturbed_matrix_equations} with $\mathbf{N} = 0$). Because there are four wave families in rotating incompressible MHD, each modified differently by changing $B_0$, when the growth rates of the individual modes cross, there appear piecewise continuous solutions. We highlight one such point in the inset in figure~\ref{fig:growth_rate_vs_B0}, where the MRI mode becomes more stable than another mode which is always stable. Since all of these piecewise discontinuities are below $\gamma=0$, they do not affect the analysis here.

The destabilizing substitution is made, and Equations \ref{eqset1} - \ref{eqset4} are rewritten such that the fluid variables are contained in a state vector 

\beq
\mathbf{V} = \left[\Psi, u_y, A, B_y\right]^\mathrm{T}.
\eeq

This yields the system of equations

\beq
\label{eq:unperturbed_matrix_equations}
 \mathcal{D}\partial_t \mathbf{V} +  \mathcal{L} \mathbf{V} + \epsilon^2\widetilde{\mathcal{G}} = \mathbf{N},
\eeq

where we leave the definition of the matrices $\mathcal{D}$, $\mathcal{L}$, and $\widetilde{\mathcal{G}}$ to Appendix \ref{app:matrices}, and the detailed form of the nonlinear vector $\mathbf{N}$ to Appendix \ref{app:nonlinear_terms}. We solve this system subject to no-slip, perfectly conducting radial boundary conditions, defined as

\beq\label{eq:wallbcs}
\Psi = \partial_x \Psi = u_y = A = \partial_x B_y = 0.
\eeq

\section{Weakly Nonlinear Analysis}\label{sec:wnl}

We conduct a formal multiple scales analysis of this system. Our perturbations are characterized in terms of fast- and slow-moving variables, that we treat as independent in order to simultaneously track the evolution of the system on two scales. The relative scalings of the fast and slow variables are chosen such that each of the temporal and spatial eigenvalues appear at the same lowest order in the linear dispersion relation (Appendix \ref{app:dispersion}). The scalings are

\beq\label{scalings}
X \equiv \epsilon x,  \, \, Y \equiv \epsilon y, \, \, Z \equiv \epsilon z, \, \, T \equiv \epsilon^2 t.
\eeq

Note that these are the same scalings as apply to Rayleigh-B\'enard convection and hydrodynamic TC flow. Our $x$ dimension, the direction of angular momentum transport, is analogous to the direction of temperature transport in the convection problem. In analogy to these problems, we posit slow variation in both $Z$ and $T$. Each operator in Equations \ref{eqset1} - \ref{eqset4} is expanded to reflect these scalings -- for instance, $\partial_z$ becomes $\partial_z + \epsilon\partial_Z$. 

The multiple scale dependencies of our solution are encoded into an ansatz for the linear MRI solution at marginality,

\beq
\label{V1_ansatz}
\mathbf{V_1} = \alpha(T, Z) \mathbb{V}_{11}(x) e^{i k_c z} + c.c. + \beta(T, Z)\mathbb{U}_{11}(x)\\
\eeq

where $\alpha(T,Z)$ is a slowly-varying amplitude and $c.c.$ denotes the complex conjugate. The $x$ dependence is contained in $\mathbb{V}_{11} = (\Psi_{11}, u_{11}, A_{11}, B_{11})^\mathbf{T}$, and must be solved subject to the radial boundary conditions. The periodic vertical boundary conditions allow us to posit the $z$ dependence, where $k_c$ is the value of the vertical wavenumber at marginality. As noted by URM07, there exists a spatially constant neutral mode solution to the $B_y$ equation, with $\mathbb{U}_{11} = (0, 0, 0, 1)^\mathbf{T}$. The amplitude $\beta(T, Z)$ encodes the slow evolution of this mode. This spatially constant mode cannot contribute to the nonlinear saturation of the MRI because all of the nonlinearities involve derivatives. The long-term evolution of $\beta(T, Z)$ is described by a simple diffusion equation that decouples from $\alpha(T, Z)$, and so we neglect it in what follows. 

The state vector is expanded in a perturbation series in orders of $\epsilon$,

\beq
\label{eq:pert_exp}
\mathbf{V} = \epsilon\mathbf{V_1} + \epsilon^2\mathbf{V_2} + \epsilon^3\mathbf{V_3} + h.o.t.
\eeq

Our perturbed system is then expressed order by order as

\begin{eqnarray}
\mathcal{O}(\epsilon):& \mathcal{L}\mathbf{V_1} + \mathcal{D}\partial_t \mathbf{V_1} = 0 \label{eq:ordere}\\
\mathcal{O}(\epsilon^2):& \mathcal{L}\mathbf{V_2} + \widetilde{\mathcal{L}}_1 \partial_Z \mathbf{V_1} = \mathbf{N_2} \label{eq:ordere2}\\
\mathcal{O}(\epsilon^3):& \mathcal{L} \mathbf{V_3} + \mathcal{D}\partial_T \mathbf{V_1}  + \widetilde{\mathcal{L}}_1\partial_Z\mathbf{V_2} \\
&+ \widetilde{\mathcal{L}}_2\partial_Z^2\mathbf{V_1} + \widetilde{\mathcal{G}}\mathbf{V_1} = \mathbf{N_3}. \label{eq:ordere3}
\end{eqnarray}

The partial differential equations that comprise Equations \ref{eq:ordere} to \ref{eq:ordere3} are solved in succession. The practical advantage of our ansatz construction (Equation \ref{V1_ansatz}) is clear: the separable x-dependence means that the radial boundary conditions are solved in only one dimension. Thus our analytical framework is able to side-step many of the resolution issues faced by multidimensional simulations. We are able to resolve even small-scale structure in the boundary layers of our domain, because we need only resolve it in one dimension. We solve the radial component of each equation using the open source pseudospectral code Dedalus. We compute the radial components on a grid of Chebyshev polynomials, as is appropriate for bounded one-dimensional domains \citep[e.g.]{Boyd:2001aa}. The nonuniform spacing of the Chebyshev grid allows us to resolve the boundary layers well on a 128-point grid.

To close the perturbation series we enforce a solvability criterion on Equation \ref{eq:ordere3} (see Appendix \ref{app:matrices}). This leads to an amplitude equation for $\alpha(T, Z)$ that governs the slow length- and timescale evolution of the system. This amplitude equation is 

\beq
\label{eq:gle}
\partial_T \alpha = b \alpha + h \partial_Z^2 \alpha - c \alpha \left|\alpha^2\right|,
\eeq

a real Ginzburg-Landau equation. The saturated solution to Equation \ref{eq:gle} is evidently $\alpha_{saturation} = \pm \sqrt{b/c}$. We plot the first order, second order, and total perturbation structure of the fluid variables in Figures \ref{fig:allorders_velocity} and \ref{fig:allorders_Bfield} with a constant $\alpha_{saturation}$. This is the Ginzburg-Landau equation that was previously found by URM07. Those authors investigated the behavior of this MRI system as a function of $\Pm$. By analyzing the system over several orders of magnitude in $\Pm$, we reproduce the URM07 result that the analytic saturation amplitude scales as $\alpha_{saturation}^2 \propto \Pm^{4/3}$ in a thin-gap geometry when $\Pm \ll 1$. 

\section{Shearing box and ambipolar diffusion}\label{sec:SBAD}

\begin{figure}
\centering
\includegraphics[width=\columnwidth]{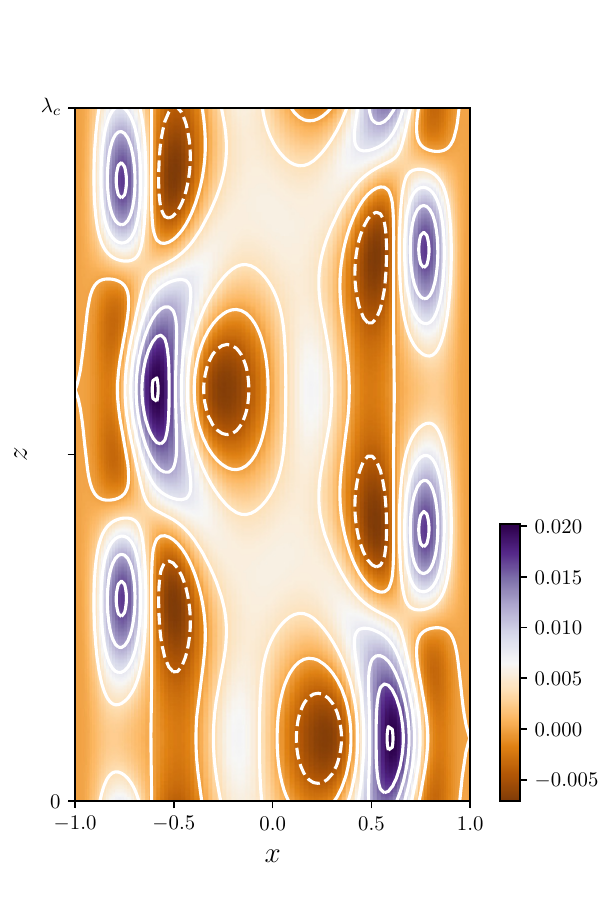}
\caption{Total (Reynolds + Maxwell) stress in the domain as predicted from the weakly nonlinear theory at $\Pm = 10^{-2}$.} \label{fig:wnl_stress}
\end{figure}

Many studies of the MRI consider the instability in a shearing box, i.e. a wall-less local approximation that is meant to represent a small section of a disk. The shearing box is the limit in which Equations \ref{eqset1} - \ref{eqset4} are subjected to shear periodic radial boundary conditions rather than Equation \ref{eq:wallbcs} \citep[e.g.][]{2008A&A...481...21R}. The periodic nature of the shearing box allows us to decompose the fluid perturbations into Fourier modes proportional to $e^{i k_x x + i k_z z}$. This makes the shearing box MRI straightforward to treat analytically. However, as noted above, the fastest-growing linear MRI modes in the shearing box are also exact solutions of the nonlinear MRI equations -- that is, $J(\hat{\psi_0}, \hat{\psi_0}) = J(\hat{\psi_0}, \nabla^2 \hat{\psi_0}) = 0$ for $\hat{\psi_0} \propto e^{i k_x x + i k_z z}$. While this may be an appealing trait for analytic simplicity, it leads to the unphysical conclusion that the fastest growing modes will never nonlinearly interact \citep{Goodman:1994ul}. This `nonlinear property' will not be satisfied for two MRI modes with nonparallel wavenumbers, but with vertically periodic boundary conditions and a vertical background magnetic field the most unstable mode has a strictly axial vertical wavenumber. Thus a formal weakly nonlinear analysis cannot be conducted, as the most unstable mode will never nonlinearly interact with itself or its complex conjugate. Similarly, we cannot analytically examine interactions between MRI channel modes and damped eigenmodes belonging to other wave families. This is analytically examined for other plasma instabilities by tracking the amplitudes of growing, marginal, and damped eigenmodes simultaneously \citep[e.g.][]{Makwana:2011}. While the shearing box approximation allows the projection of the perturbed MRI equations into the basis set of linear eigenmodes, nonlinear coupling between modes will remain zero.

\begin{figure*}
\centering
\includegraphics[width=\textwidth]{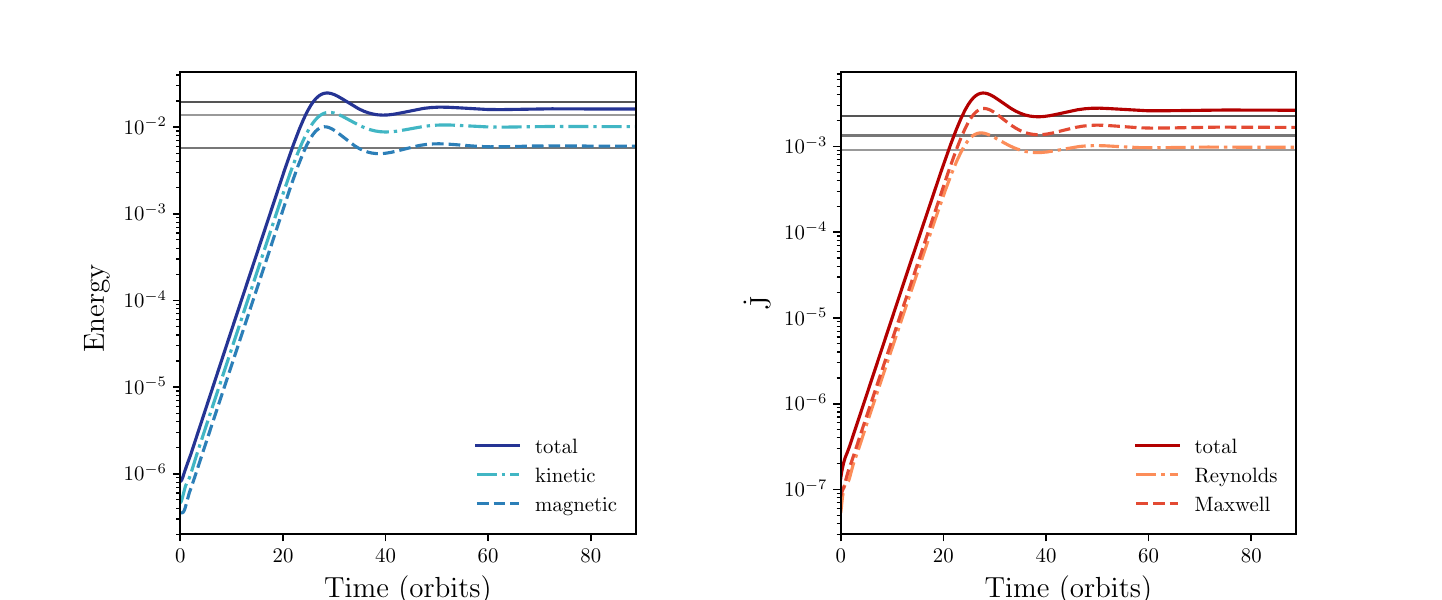}
\caption{Average energy (left) and angular momentum transport (right) in the total, kinetic, and magnetic components of simulation data as a function of time. Gray lines show the weakly nonlinear theory values for each quantity.} \label{fig:sim_theory_energy_jdot}
\end{figure*}

The nonlinear property of primary MRI modes in the shearing box motivates the addition of radial boundaries, such that the nonlinear evolution of the weakly nonlinear MRI can be properly considered. It also raises the question of whether some additional nonlinear mechanism can be introduced such that the fastest-growing modes are no longer nonlinear solutions to the shearing box equations. It has already been shown that the Hall effect does not negate the nonlinear property of primary MRI modes \citep{Kunz:2013}. However, it seems to have been overlooked in the literature that these linear modes are \textit{not} solutions of the nonlinear ambipolar diffusion term, which is proportional to

\beq
\nabla \times ((\mathbf{J} \times \mathbf{B}) \times \mathbf{B}). 
\eeq

Furthermore, the radial wavenumber of the fastest-growing linear MRI mode in a shearing box with ambipolar diffusion is nonzero when a constant azimuthal background field is considered in addition to an axial one \citep{Kunz:2004ib}. This means that, in the presence of ambipolar diffusion, we can derive the weakly nonlinear envelope equation for the MRI in the shearing box. Ambipolar diffusion adds both linear and nonlinear terms to Equations \ref{eq:ordere} to \ref{eq:ordere3}, but does not change their $Z$ or $T$ dependence. The constant azimuthal background field component does not contribute to any other terms in the local MRI equations. Thus, the slow-scale evolution of the MRI in a shearing box with ambipolar diffusion is also governed by a Ginzburg-Landau equation.

The Ginzburg-Landau form of the amplitude equation can be found in any system with Euclidean symmetry and a quadratic maximum in growth rate with respect to the wavenumber \citep{Hoyle:2006}. In this case, the Euclidean symmetry comes from axisymmetry in the $x$-$z$ plane, and the quadratic maximum is a consequence of the linear dispersion relation given in Appendix~\ref{app:dispersion}. In Paper II, we show that the same symmetry occurs in the axisymmetric global geometry as well.
The Ginzburg-Landau equation arises due to symmetries in the local MRI equations, irrespective of the boundary conditions to which they are subjected. This means that the local MRI is able to saturate via nonlinear mode interaction so long as the primary MRI modes are not exact solutions of the nonlinear terms. This can be achieved by considering the effects of ambipolar diffusion when the boundary conditions are shear periodic, or by enforcing wall-like radial boundary conditions. Both constructions require the most unstable mode to have nonconstant radial structure. Physically, this radial variation impedes the free exchange of angular momentum facilitated by the uniform stretching of channel modes. 

\section{Direct Numerical Simulation}\label{sec:simulations}

Here, we make a preliminary test of our weakly nonlinear theory by comparing it to direct numerical simulation. Using Dedalus, we  solved the full, nonlinear equations \ref{eqset1} - \ref{eqset4} with all parameters ($\reym$, $Q$) equal to their critical values except the background magnetic field, which we set to $B_0 = 1 + \epsilon^2$. We thus drive the system MRI unstable in the same way as in our theory. The computational requirements of low $\Pm$ simulations are quite intense in both time and space. Despite being virtually smooth, the solutions require a resolution of $192 \times 1536$ grid points at $\Pm = 10^{-2}$. Because the system has such a small growth rate, it takes hundreds of orbits for the system to reach saturation, as compared to the few orbits typical of high $\reym$ simulations \citep[e.g.][]{2007MNRAS.378.1471L}. As a result, we make our comparison at $\Pm = 10^{-2}$, which provides a good tradeoff between probing relatively low $\Pm$ while keeping the computational time for these exploratory simulations modest.

We initialize the runs with the linear eigenvectors of the MRI unstable mode (also computed by Dedalus; see section~\ref{sec:wnl}) multiplied by an initial amplitude $A_0 = 10^{-3}$. Doing so requires considerably less run time, as the MRI unstable mode starts growing immediately from $A_0$. By contrast, initializing random noise in $\psi$ with amplitude $A_0$ would give the unstable mode a much smaller amplitude. Nevertheless, we have confirmed that simulations with eigenvector initial conditions have similar evolutions to those with noise initial conditions once each enter linear growth.

\begin{figure}
\centering
\includegraphics[width=\columnwidth]{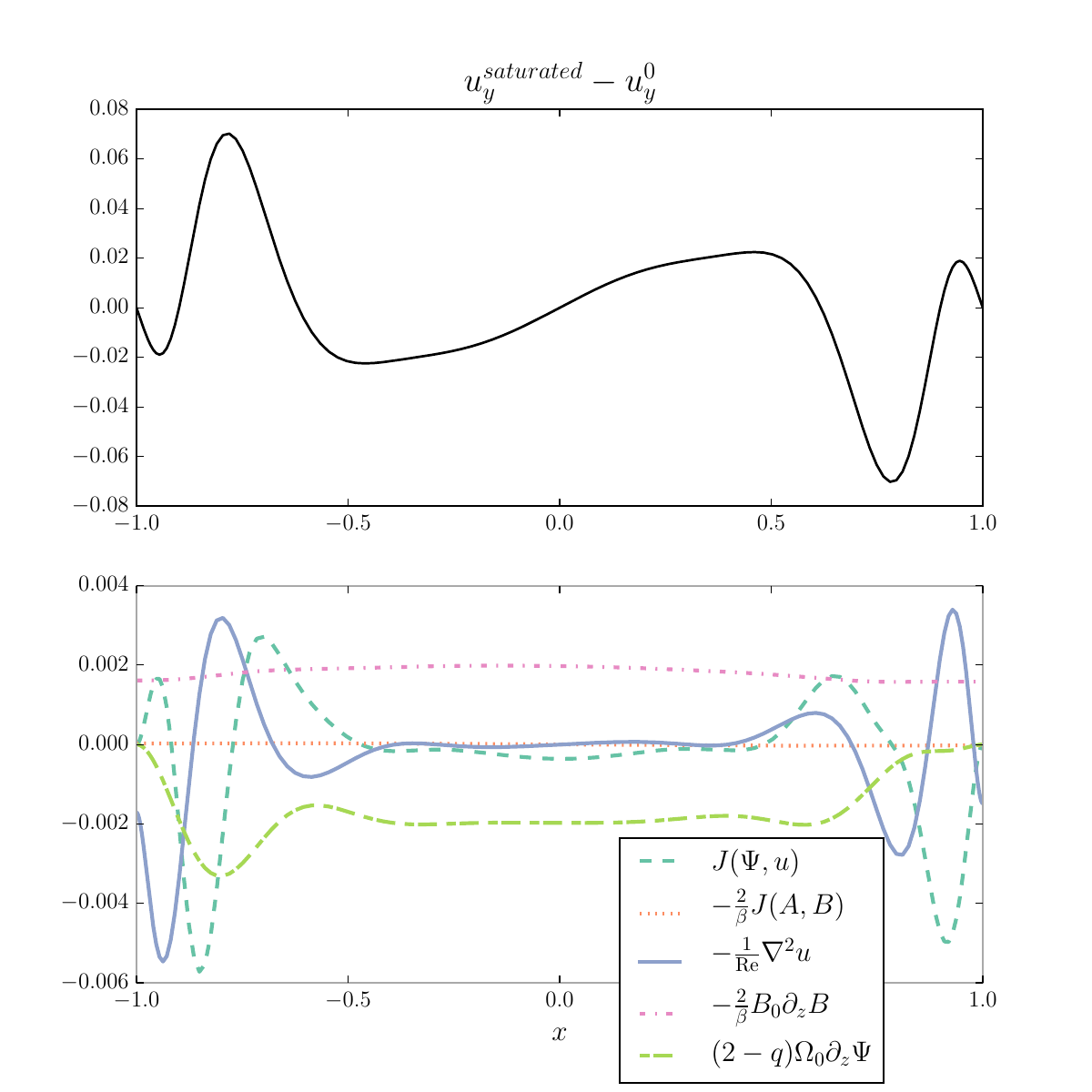}
\caption{Top panel: radial profile for $u_y^{saturated} - u_y^0$, the difference between the saturated azimuthal velocity profile and the initial azimuthal velocity profile (TC flow). Bottom panel: each term in the steady state force balance (Equation \ref{eqset2} with $\partial_t u_{y} = 0$). Saturated quantities are computed with $\alpha_{saturation} = \sqrt{b/c}$ and $\epsilon = 0.5$. The saturated state shows reduced shear in the bulk of the flow, outside of the boundary layers.} \label{fig:sat_comp_vel}
\end{figure}

\begin{figure}
\centering
\includegraphics[width=\columnwidth]{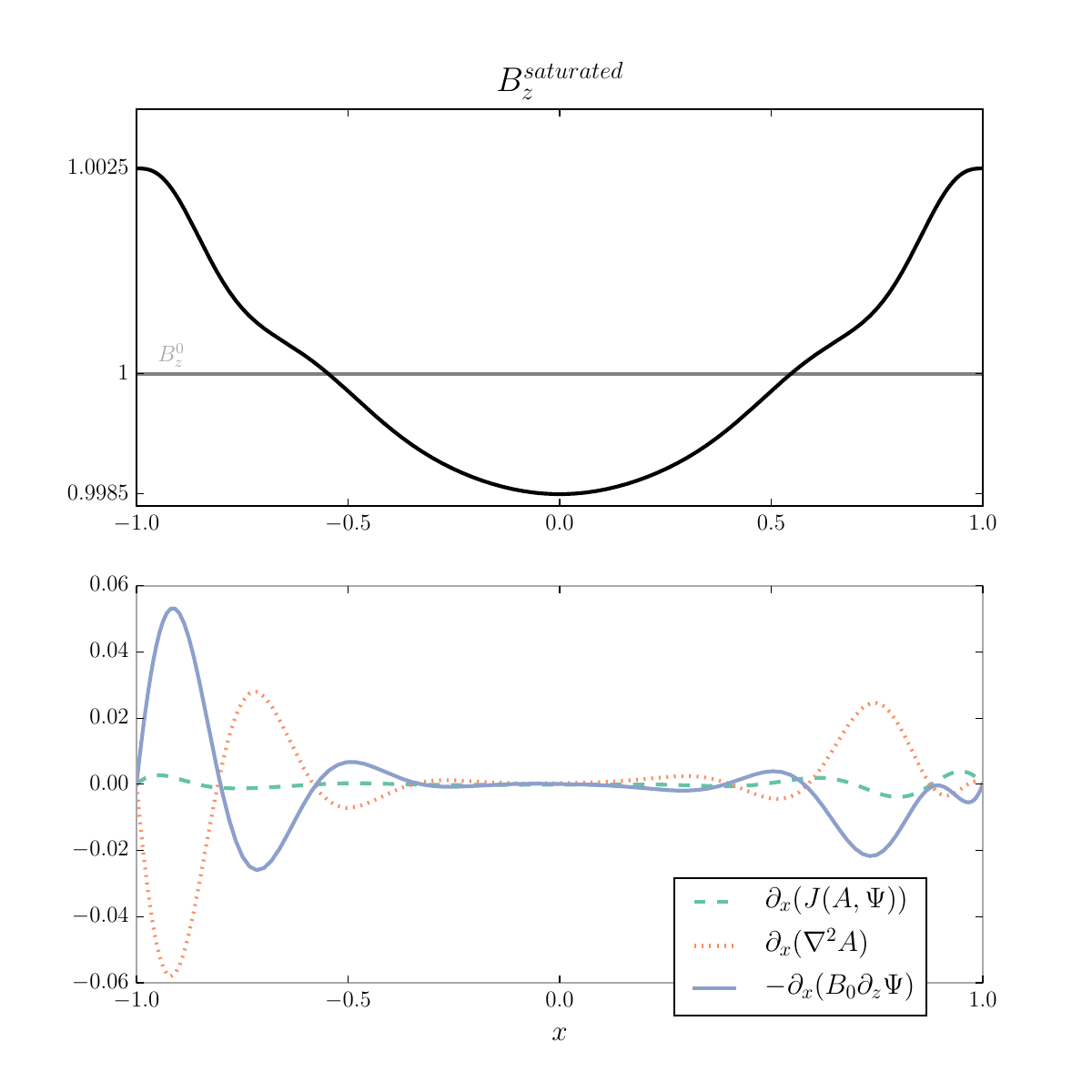}
\caption{Top panel: radial profile for $B_z^{saturated}$, the saturated vertical magnetic field (black line). $B_z^0 = 1$ is the constant background magnetic field (gray line). Bottom panel: each term in the steady state inductive balance ($\partial_x$ of Equation \ref{eqset4} with $\partial_t B_y = 0$). Saturated quantities are computed with $\alpha_{saturation} = \sqrt{b/c}$ and $\epsilon = 0.5$. The saturated field is pushed to the radial domain boundaries.}\label{fig:sat_comp_Bfield}
\end{figure}

We analyze the average energy and angular momentum transport in the simulation domain (Figure \ref{fig:sim_theory_energy_jdot}). The saturation amplitude predicted by the weakly nonlinear theory depends on the choice of normalization of the linear eigenvectors. The eigenvectors of the linear problem are only determined up to an arbitrary normalization, and the nonlinear coefficient of the Ginzburg-Landau equation is sensitive to this normalization. The undetermined factor is typically assigned by comparison with direct numerical simulation or laboratory experiment \citep[e.g.][]{Deyirmenjian:1997,Recktenwald:1993}. Here we determine the constant by requiring that the maximum amplitude of $B_y$ be equal in both theory and simulation. With this normalization choice all plotted quantities agree to within $\sim 25\%$. The theory and simulation are thus in reasonably good agreement considering that the weakly nonlinear theory applies rigorously to a channel of infinite height, while the simulation was carried out in a box with a vertical extent of only two critical wavelengths. We defer further comparisons between simulation and theory, including an analysis of the effect of the box height on the simulated flow, to future work.

\section{Discussion}
\label{sec:discussion}

Here our focus is on a physical description of the saturation mechanism. Figure~\ref{fig:sat_comp_vel} shows saturated radial profiles of $u_0 - u_{y} = -q \Omega_0 x - u_{y}$ and each term in the steady state force balance (i.e. Equation~\ref{eqset2} with $\partial_t u_{y} = 0$). In the bulk of the fluid away from the boundary layers, the saturated state shows reduced shear, with little diffusive contribution. This demonstrates that even in a case where diffusive effects are important, the bulk of the fluid saturates by balancing shear and magnetic tension. As discussed at length in \citet{Vasil:2015}, when diffusive effects are not important, it is impossible to rearrange momentum without also rearranging the magnetic field. The \citet{Vasil:2015} model demonstrates saturation without diffusive effects; our results show that outside of the boundary layers, a simultaneous rearrangement of momentum and field occurs. In the boundary layers, the nonlinear advection balances viscous dissipation. 

\begin{figure*}[h!]
\centering
\includegraphics[width=\textwidth]{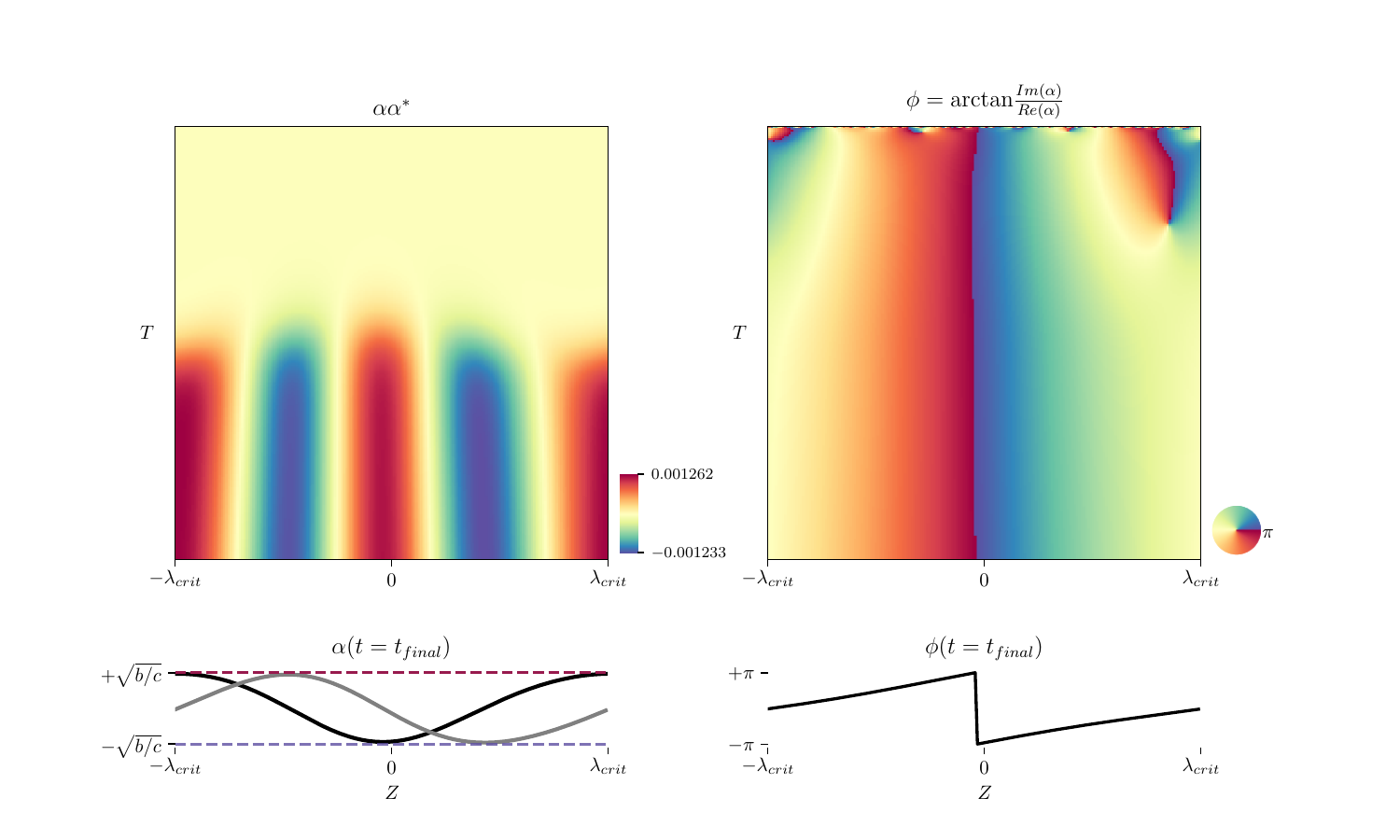}
\caption{Evolution of the Ginzburg-Landau amplitude equation (Equation \ref{eq:gle}) on a Fourier $Z$ domain of length $2 \lambda_{crit}$, where $\lambda_{crit} = 2\pi/k_c$ is the critical wavelength of the system. Top left panel shows the evolution of the amplitude observable $\alpha \alpha^*$ on the full $Z$ domain as a function of time $T$. Bottom left panel shows the amplitude $alpha$ at the final timestep shown, where the black line is the real part $\mathrm{Re}\{\alpha(t = t_{final})\}$ and the gray line is the imaginary part $\mathrm{Im}\{\alpha(t = t_{final})\}$. The final amplitude is bounded by the analytic saturation amplitude $\alpha_{saturation} = \pm \sqrt{b/c}$. Top right panel shows the evolution of the phase angle $\phi = \mathrm{arctan} (\mathrm{Im}(\alpha)/\mathrm{Re}(\alpha))$ on the same domain. Bottom panel shows the phase angle as a function of $Z$ for the final timestep. Note that the phase angle is wrapped on a $2\pi$ domain, such that $\pi$ = $-\pi$, as indicated by the circular colorbar.}\label{fig:IVP_2lambdacrit}

\centering
\includegraphics[width=\textwidth]{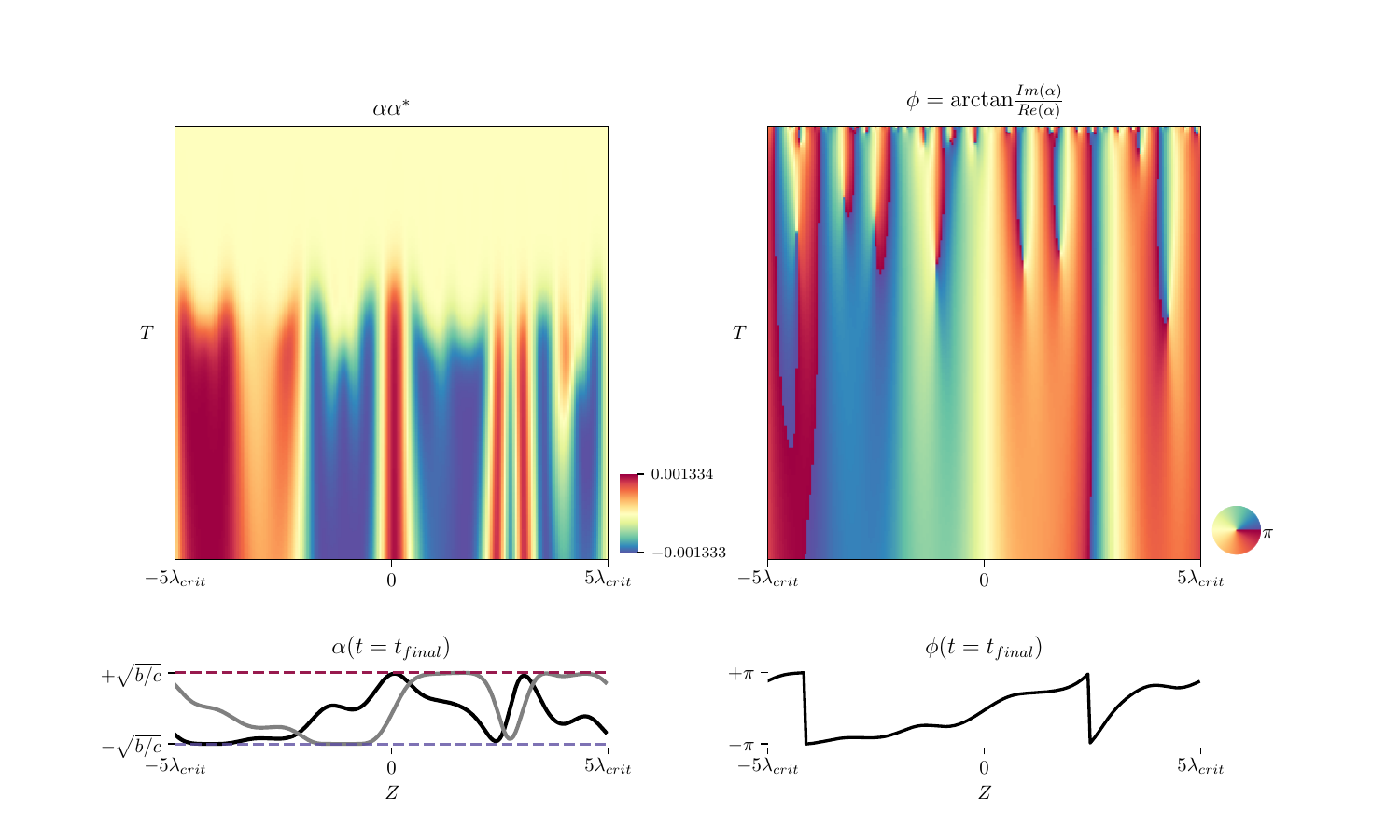}
\caption{As in Figure \ref{fig:IVP_2lambdacrit} but for a $Z$ domain of length $10 \lambda_{crit}$.}\label{fig:IVP_10lambdacrit}
\end{figure*}

Figure~\ref{fig:sat_comp_Bfield} shows $B_z$ and the terms corresponding to steady state inductive balance ($\partial_x$ of Equation~\ref{eqset4} with $\partial_t B_{y} = 0$). Here, the instability acts to push the magnetic field toward the boundaries in both the bulk and the boundary layers. 
The radial average of the saturated $B_z$ is $B_0$, i.e. $B_z$ is marginally stable.
 \citet{Ebrahimi:2009ey} considered the saturation of a single, strongly super-critical MRI mode allowed to interact nonlinearly only with itself and the mean. They considered two important cases, one in which the mean flow was forced to remain at its initial, quasi-Keplerian state for all time, and one in which the background flow was allowed to evolve. This is a crucial difference between the shearing box and our narrow-gap TC flow: perturbations in our simulation can adjust the background flow, whereas in a shearing box, the shear periodicity forbids perturbations from affecting the mean flow. In the case with a freely evolving background flow, \citet{Ebrahimi:2009ey} found a saturated state quite similar to ours: field pushed to the boundaries, and a reduction in shear in the bulk of the flow. Their flows have less pronounced boundary layers, likely because of their much larger $\Pm = 0.1 -1$. 

In the high $\reye$ and  $\reym$  limit, \citet{Vasil:2015} derives an amplitude equation considerably different than the one found here. By averaging in the $z$ direction, the author computes a mean-field equation with striking similarity to the buckling of an elastic beam under load. The most salient feature of this equation is its \emph{non-local} character. Unlike the present work, which focuses on Keplerian rotation profiles with $q = 3/2$ with a critical background magnetic field strength, \citet{Vasil:2015} focuses on a fixed field strength and a weakly destabilized shear profile. These differences are minor, however: the destabilizing parameter $\epsilon$ enters the analysis in the same quadratic proportion. Whether and how \citet{Vasil:2015}'s amplitude equation is equivalent to our own in the limit of dynamically important resistive and viscous effects is beyond the scope of this work. Nevertheless, the author identifies the nonlinear term responsible for saturation as consisting of flux and field transport and notes these are the only mechanisms able to produce saturation. Our results likewise demonstrate a combination of flux and field transport in the comparable region of our domain. This suggests that despite our formulation displaying different saturation dynamics (Ginzburg-Landau in our case; a network of coupled Duffing oscillators in \citeauthor{Vasil:2015} \citeyear{Vasil:2015}), there may indeed be an underlying unification. 

The real Ginzburg-Landau equation describes the amplitude behavior of our system close to threshold. Although the form of the equation is generic to many systems, its coefficients depend on the specific physics of our system and govern its detailed evolution (see Appendix \ref{app:matrices}). We simulate the evolution of the MRI amplitude equation by solving Equation \ref{eq:gle} on a Fourier basis in $Z$ using Dedalus. We initialize uniform random noise of amplitude $-10^{-3}$ to $+10^{-3}$ in $Z$, and timestep the system using a four-stage, third-order Runge-Kutta integrator. We evolve the system for $100 \Omega_0^{-1}$ in timesteps of $0.02 \Omega_0^{-1}$. Results are shown in Figures \ref{fig:IVP_2lambdacrit} and \ref{fig:IVP_10lambdacrit}, where the amplitude and phase structure over the vertical domain is plotted for every $20$ timesteps. The system quickly organizes itself into rolls in $Z$ bounded by the analytic saturation amplitude $\alpha_s = \sqrt{b/c}$. The specific geometry depends on the number of critical wavelengths $\lambda_{crit} = 2\pi/k_c$ that are initialized in $Z$. Figure \ref{fig:IVP_2lambdacrit} shows that a system with a height equal to two critical wavelengths will be modulated by simple rolls of sinusoidal amplitude. The saturation amplitude pattern becomes more complicated when more modes are allowed to interact. Figure \ref{fig:IVP_10lambdacrit} shows the evolution of a system of height $10 \lambda_{crit}$. While still bounded by $\alpha_s$, the saturation amplitude exhibits a nonlinear phase geometry due to the nonlinear interaction of modes in $Z$. 

The weakly nonlinear theory predicts that the amplitude of the system is bounded by the saturation amplitude $\alpha_s = \sqrt{b/c}$, where $b$ and $c$ are coefficients corresponding to the linear growth term and nonlinear term of the Ginzburg-Landau equation, respectively. The coefficient $b$ comes from the interaction between the background magnetic field and the linear MRI solution. The coefficient $c$ describes the third-order nonlinear interaction between terms in the perturbation series. Physically, we see that the saturation amplitude is controlled by the strength of the mode interaction within our finite band of unstable modes. We stress that while the third-order nonlinear terms in the walled TC flow are strongly influenced by the boundary layers, this is not generically true of the MRI system. Indeed, in the shearing box MRI with ambipolar diffusion (the case sketched out in Section \ref{sec:SBAD}), boundary layers are impossible in the shear periodic flow. In this case the third-order nonlinear behavior of the system includes three-mode interactions from the cubic nonlinearity in the ambipolar diffusion term.

Figure~\ref{fig:wnl_stress} shows the total stress $u_x u_y - \Co B_x B_y$ for the $\Pm = 10^{-2}$ model with $\epsilon = 0.5$. The stress shows significantly more structure throughout the domain than the variables $u_x$, $u_y$, $B_x$ and $B_y$ that comprise it, demonstrating that a non-trivial correlation exists even in the weakly non-linear state. As in simulations at higher $\reym$, figure~\ref{fig:sim_theory_energy_jdot} shows that the Maxwell stress dominates over the Reynolds stresses even though the kinetic energy significantly exceeds the magnetic energy. 

\section{Conclusion}
\label{sec:conclusion}

In this paper we construct a weakly nonlinear analysis of the MRI using multiple scales analysis, leading to a real Ginzburg-Landau equation for the nonlinear amplitude, confirming the previous results of \citet{Umurhan:2007hs}. We also confirm their results for the scaling of the analytic saturation amplitude with $\Pm$. We extend their results by constructing a detailed force and inductive balance for the saturated $u_y$ and $B_z$ components. In doing so, we find the saturated state is a complex balance in which reduction of shear and amplification and redistribution of $B_z$ combine to saturate the instability. We perform numerical simulations of the amplitude equation and a direct numerical simulation of the MRI system. Using the former, we demonstrate that complex patterns can organize the flow on long length scales $Z$, though the maximum magnitude of the amplitude $\alpha$ is well predicted by the steady state solution. The latter show that there is rough agreement for both total energy and average angular momentum transport between the weakly nonlinear theory and simulation for a representative case at $\Pm = 10^{-2}$. We defer a full comparison between theory and simulation to later work. We describe the application of shear-periodic boundary conditions to the local MRI and find that with the inclusion of certain nonideal physical effects, namely ambipolar diffusion, our theory points to a new saturation avenue for the MRI in a shearing box. In Paper II, we make use of the techniques developed here to extend the weakly nonlinear analysis of the MRI to a full cylindrical geometry appropriate for a Taylor-Couette experiment. 

\section{Acknowledgments}
S.E.C. was supported by a National Science Foundation Graduate Research Fellowship under grant No. DGE-16-44869. J.S.O. acknowledges support from NASA grant NNX16AC92G. We thank the anonymous referee for thoughtful comments that greatly improved the manuscript. We also thank Mordecai-Mark Mac Low, Jeremy Goodman, John Krommes, Geoff Vasil, and Ellen Zweibel for useful discussion. \software{Dedalus: http://dedalus-project.org/}

\clearpage
\appendix

\section{Detailed Equations}\label{app:matrices}

Here we detail the perturbation analysis described in Section \ref{sec:wnl}. The perturbation series is described by Equations \ref{eq:ordere} - \ref{eq:ordere3}, where 

\beq
\mathcal{L} = \mathcal{L}_0 + \mathcal{L}_1 \partial_z + \mathcal{L}_2 \partial_z^2 + \mathcal{L}_3 \partial_z^3 + \mathcal{L}_4 \partial_z^4,
\eeq

\beq
\widetilde{\mathcal{L}}_1 =  \mathcal{L}_1 + 2\mathcal{L}_2\partial_z + 3\mathcal{L}_3\partial_z^2 + 4\mathcal{L}_4\partial_z^3
\eeq

\beq
\widetilde{\mathcal{L}}_2 = \mathcal{L}_2 + 3\mathcal{L}_3\partial_z + 6\mathcal{L}_4\partial_z^2
\eeq

\beq
\widetilde{\mathcal{G}} = \mathcal{G} \partial_z + \mathcal{L}_3 \partial_z^3,
\eeq

and the constituent matrices are defined as 

\beq
\mathcal{D} = \left[\begin{matrix}
\nabla^2 & 0 & 0 & 0 \\
0 & 1& 0 & 0 \\
0 & 0 & 1 & 0\\
0 & 0 & 0 & 1 \\
\end{matrix}\right]
\eeq

\beq
\mathcal{L}_0 = \left[\begin{matrix}
-\frac{1}{\reye}\partial_x^4 & 0 & 0 & 0 \\
0 & -\frac{1}{\reye}\partial_x^2 & 0 &0 \\
0 & 0 & -\frac{1}{\reym}\partial_x^2 & 0 \\
0 & 0 & 0 & -\frac{1}{\reym}\partial_x^2 \\ \end{matrix}\right]
\eeq

\beq
\mathcal{L}_1 = \left[\begin{matrix}
0 & -2 & -\Co\partial_x^2 & 0 \\
(2-q)\Omega_0 & 0 & 0 & -\Co \\
-1 & 0 & 0 & 0 \\
0 & -1 & q\Omega_0 & 0 \\ \end{matrix}\right] 
\eeq

\beq
\mathcal{L}_2 = \left[\begin{matrix}
-2\frac{1}{\reye} \partial_x^2 & 0 & 0 & 0 \\
0 & -\frac{1}{\reye} & 0 & 0 \\
0 & 0 & -\frac{1}{\reym} & 0 \\
0 & 0 & 0 & -\frac{1}{\reym} \\ \end{matrix}\right]
\eeq

\beq
\mathcal{L}_3 = \left[\begin{matrix}
0 & 0 & -\Co & 0 \\
0 & 0 & 0 & 0 \\
0 & 0 & 0 & 0 \\
0 & 0 & 0 & 0 \\ \end{matrix} \right]
\eeq

\beq
\mathcal{L}_4 = \left[\begin{matrix}
-\frac{1}{\reye} & 0 & 0 & 0 \\
0 & 0 & 0 & 0 \\
0 & 0 & 0 & 0 \\
0 & 0 & 0 & 0 \\ \end{matrix}\right] 
\eeq

\beq
\mathcal{G} = \left[\begin{matrix}
0 & 0 & -\Co\partial_x^2 & 0 \\
0 & 0 & 0 & -\Co \\
-1 & 0 & 0 & 0 \\
0 & -1 & 0 & 0 \\ \end{matrix} \right]
\eeq

Once perturbed, the system is solved for successive orders of $\epsilon$ (Equations \ref{eq:ordere} - \ref{eq:ordere3}). $\mathcal{O}(\epsilon)$ is the linear system. At $\mathcal{O}(\epsilon^2)$, first-order MRI modes nonlinearly interact with themselves and with their complex conjugates, and so the term $\mathbf{N}_2$ in Equation \ref{eq:ordere2} has the form

\beq
\mathbf{N}_2 = |\alpha|^2 \mathbf{N}_{20} + \alpha^2 \mathbf{N}_{22} e^{2 i k_c z}
\eeq

(see Appendix \ref{app:nonlinear_terms} for the full form of $\mathbf{N}_{20}$ and $\mathbf{N}_{22}$). 

Note that, following the notation of \citet{Umurhan:2007hs}, the subscripts refer to $\epsilon$ order, $z$ order, successively, such that $\mathbf{N}_{22}$ is the second-order nonlinear term which corresponds to $e^{2 i k_c z}$ z-dependence.  

Equation \ref{eq:ordere2} is solved as three separate systems of equations, one for each possible $z$ resonance: 

\begin{align}
\mathcal{L}\mathbf{V}_{20} & = \mathbf{N}_{20}\\
\mathcal{L}\mathbf{V}_{21} & = - \widetilde{\mathcal{L}}_1 \partial_Z \mathbf{V}_{11} \\
\mathcal{L}\mathbf{V}_{22} & = \mathbf{N}_{22}
\end{align}

Finally, at $\mathcal{O}(\epsilon^3)$ we eliminate secular terms to close the system. Secular terms are terms which are resonant with the solution to the homogenous linear equation (Equation \ref{eq:ordere}), and which cause the higher-order solutions to grow without bound. The solvability criterion we enforce to eliminate these terms is the vanishing of the inner product of the solution to the adjoint linear homogenous equation $\mathcal{L}^\dagger \mathbf{V}^\dagger = 0$ with the nonhomogenous terms in Equation \ref{eq:ordere3}, namely

\beq\label{eq:innerprod_gle}
\langle \mathbb{V}^\dagger | \mathcal{D} \mathbb{V}_{11} \rangle \partial_T \alpha + \langle  \mathbb{V}^\dagger | \widetilde{\mathcal{G}} \mathbb{V}_{11} \rangle \alpha +  \langle \mathbb{V}^\dagger | \widetilde{\mathcal{L}}_1 \mathbb{V}_{21} + \widetilde{\mathcal{L}}_2 \mathbb{V}_{11} \rangle \partial_Z^2 \alpha = \langle \mathbb{V}^\dagger | \mathbf{N}_{31} \rangle \alpha |\alpha|^2.
\eeq

This solvability criterion derives from a corollary to the Fredholm Alternative (see Paper II for a formal definition). 

Equation \ref{eq:innerprod_gle} can be rewritten as Equation \ref{eq:gle}, the Ginzburg-Landau equation, where the coefficients are 

\beq
b = \langle  \mathbb{V}^\dagger | \widetilde{\mathcal{G}} \mathbb{V}_{11} \rangle / \langle \mathbb{V}^\dagger | \mathcal{D} \mathbb{V}_{11} \rangle,
\eeq

\beq
h = \langle \mathbb{V}^\dagger | \widetilde{\mathcal{L}}_1 \mathbb{V}_{21} + \widetilde{\mathcal{L}}_2 \mathbb{V}_{11} \rangle / \langle \mathbb{V}^\dagger | \mathcal{D} \mathbb{V}_{11} \rangle,
\eeq

and

\beq
c = \langle \mathbb{V}^\dagger | \mathbf{N}_{31} \rangle / \langle \mathbb{V}^\dagger | \mathcal{D} \mathbb{V}_{11} \rangle. 
\eeq

We define the adjoint operator $\mathcal{L}^\dagger$ and solution $\mathbf{V}^\dagger$ as 

\beq
\langle \mathbf{V^\dagger} | \mathcal{L} \mathbf{V} \rangle = \langle \mathcal{L}^\dagger \mathbf{V}^\dagger | \mathbf{V} \rangle,
\eeq

where the inner product is defined as 

\beq\label{eq:inner_product_def}
\langle \mathbf{V^\dagger} | \mathcal{L} \mathbf{V} \rangle = \frac{k_c}{2\pi} \int_{-\pi/k_c}^{\pi/k_c} \int_{x_1}^{x_2} \mathbf{V}^{\dagger*} \cdot \mathcal{L} \mathbf{V} \, \mathrm{d} x \mathrm{d} z.
\eeq

The solution to the adjoint homogenous equation has the form 

\beq
\mathbf{V^\dagger} = \mathbb{V}^\dagger(x) e^{i k_c z} + c.c.
\eeq

As noted by URM07, a second amplitude equation for a spatially constant azimuthal magnetic field mode arises from the terms in the $\mathcal{O}(\epsilon^3)$ equation which contain no $z$ dependence. This is a diffusion equation, so the neutral mode simply decays away.

\section{Expansion of Nonlinear Terms}\label{app:nonlinear_terms}

At each order in our perturbation series, lower-order MRI modes nonlinearly interact. Thus there is a nonlinear term contribution at $\mathcal{O}(\epsilon^2)$ and $\mathcal{O}(\epsilon^3)$. Here we detail the form of these nonlinear terms. 

The overall nonlinear contribution to our system, written as a vector $\mathbf{N}$ in Equation \ref{eq:unperturbed_matrix_equations}, is

\beq
\mathbf{N} = \epsilon^2\mathbf{N_2} \, + \, \epsilon^3\mathbf{N_3} \, + \, \mathcal{O}(\epsilon^4)
\eeq

where

\beq
N_2^{(\Psi)} = J(\Psi_1, \nabla^2 \Psi_1) \,-\, \Co J(A_1, \nabla^2 A_1) 
\eeq

\beq
N_2^{(u)} = J(\Psi_1, u_1) \, - \, \Co J(A_1, B_1) 
\eeq

\beq
N_2^{(A)} = - J(A_1, \Psi_1) 
\eeq

\beq
N_2^{(B)} =  J(\Psi_1, B_1) \, - \, J(A_1, u_1)
\eeq

and

\beq
\begin{split}
N_3^{(\Psi)} & = J(\Psi_1, \nabla^2\Psi_2) \, - \, \Co J(A_1, \nabla^2 A_2) \,+\, J(\Psi_2, \nabla^2\Psi_1) - \, \Co J(A_2, \nabla^2 A_1) \,+\, 2 J(\Psi_1, \partial_z\partial_Z \Psi_1) \, \\
& - \, 2 \Co J(A_1, \partial_z\partial_Z A_1) \,+\, \widetilde{J}(\Psi_1, \nabla^2 \Psi_1) \, - \,  \Co \widetilde{J}(A_1, \nabla^2 A_1)\\
\end{split}
\eeq

\beq
N_3^{(u)} = J(\Psi_1, u_2) \, + \, J(\Psi_2, u_1) \, + \, \widetilde{J}(\Psi_1, u_1) \, - \, \Co J(A_1, B_2) \, - \, \Co J(A_2, B_1) \, - \, \Co \widetilde{J}(A_1, B_1)
\eeq

\beq
N_3^{(A)} = - J(A_1, \Psi_2) \, - \, J(A_2, \Psi_1) \, - \, \widetilde{J}(A_1, \Psi_1)
\eeq

\beq
N_3^{(B)} = \, J(\Psi_1, B_2) \, + \, J(\Psi_2, B_1) \, + \, \widetilde{J}(\Psi_1, B_1) \, - \, J(A_1, u_2) \, - \,  J(A_2, u_1) \, - \, \widetilde{J}(A_1, u_1).\\
\eeq

$\mathbf{N_2}$ and $\mathbf{N_3}$ expand to become

\beq
\mathbf{N_2} = \alpha^2\mathbb{N}_{22} e^{i2 k_c z} + \left|\alpha\right|^2 \mathbb{N}_{20} + c.c.
\eeq

and

\beq
\mathbf{N_3} = \alpha^3 \mathbb{N}_{33} e^{i 3 k_c z} + \alpha\partial_Z\alpha \mathbb{N}_{32} e^{i 2 k_c z} + \alpha\left|\alpha\right|^2 \mathbb{N}_{31} e^{i k_c z} + \alpha \partial_Z  \beta \mathbb{\widetilde{N}}_{31} e^{i k_c z} + \alpha^*\partial_Z \alpha \mathbb{N}_{30} + c.c.
\eeq

The second order nonlinear terms are \\

\beq
\begin{split}
N_{22}^{(\Psi)} = & \, i k_c \Psi_{11} \cdot \left(\partial_x^3 \Psi_{11} - k_c^2 \partial_x \Psi_{11}\right) - \partial_x \Psi_{11} \cdot \left(i k_c \partial_x^2 \Psi_{11} - i k_c^3 \Psi_{11}\right) \\
& + \Co \partial_x A_{11} \cdot \left(i k_c \partial_x^2 A_{11} - i k_c^3 A_{11}\right) - \Co i k_c A_{11} \cdot \left(\partial_x^3 A_{11} - k_c^2 \partial_x A_{11}\right)
\end{split}
\eeq

\beq
\begin{split}
N_{22}^{(u)} = & \, i k_c \Psi_{11} \cdot \partial_x u_{11} - \partial_x \Psi_{11} \cdot i k_c u_{11} - \Co i k_c A_{11} \cdot \partial_x B_{11} + \Co \partial_x A_{11} \cdot i k_c B_{11}
\end{split}
\eeq

\beq
N_{22}^{(A)} = - i k_c A_{11} \cdot \partial_x \Psi_{11} + \partial_x A_{11} \cdot i k_c \Psi_{11}
\eeq

\beq
N_{22}^{(B)} = i k_c \Psi_{11} \cdot \partial_x B_{11} - \partial_x \Psi_{11} \cdot i k_c B_{11} - i k_c A_{11} \cdot \partial_x u_{11} + \partial_x A_{11} \cdot i k_c u_{11}
\eeq

\beq
\begin{split}
N_{20}^{(\Psi)} = & \, i k_c \Psi_{11} \cdot \left(\partial_x^3 \Psi_{11}^* - k_c^2 \partial_x \Psi_{11}^*\right) - \partial_x \Psi_{11} \cdot \left(i k_c^3 \Psi_{11}^* - i k_c \partial_x^2 \Psi_{11}^*\right) \\
& + \Co \partial_x A_{11} \cdot \left(i k_c^3 A_{11}^* - i k_c \partial_x^2 A_{11}^* \right) - \Co i k_c A_{11} \cdot \left(\partial_x^3 A_{11}^* - k_c^2 \partial_x A_{11}^*\right)
\end{split}
\eeq

\beq
N_{20}^{(u)} = i k_c \Psi_{11} \cdot \partial_x u_{11}^* + \partial_x \Psi_{11} \cdot i k_c u_{11}^* - \Co i k_c A_{11} \cdot \partial_x B_{11}^* - \Co \partial_x A_{11} \cdot i k_c B_{11}^*
\eeq

\beq
N_{20}^{(A)} = - i k_c A_{11} \cdot \partial_x \Psi_{11}^* - \partial_x A_{11} \cdot i k_c \Psi_{11}^*
\eeq

\beq
N_{20}^{(B)} = i k_c \Psi_{11} \cdot \partial_x B_{11}^* + \partial_x \Psi_{11} \cdot i k_c B_{11}^* - i k_c A_{11} \cdot \partial_x u_{11}^* - \partial_x A_{11} \cdot i k_c u_{11}^*
\eeq \\

and the third order nonlinear terms become \\

\beq
\begin{split}
N_{31}^{(\Psi)} = &  \, i k_c \left(\Psi_{11} \cdot \partial_x^3 \Psi_{20}\right) + i k_c \left(\Psi_{11} \cdot \partial_x^3\Psi_{20}^*\right) - i k_c \left(\Psi_{11}^* \cdot \partial_x^3 \Psi_{22}\right) - i 2 k_c \left(\partial_x \Psi_{11}^* \cdot \partial_x^2 \Psi_{22}\right) \\
& + i 8 k_c^3 \left(\partial_x\Psi_{11}^* \cdot \Psi_{22}\right) + i 4 k_c^3 \left(\Psi_{11}^* \cdot \partial_x \Psi_{22}\right) + \Co \left[ - i k_c \left(A_{11} \cdot \partial^3 A_{20}\right) - i k_c \left(A_{11} \cdot \partial_x^3 A_{20}^*\right) \right]\\
& + \Co \left[i k_c \left(A_{11}^* \cdot \partial_x^3 A_{22}\right) + i 2 k_c \left(\partial_x A_{11}^* \cdot \partial_x^2 A_{22}\right) - i 8 k_c^3 \left(\partial_x A_{11}^* \cdot A_{22}\right) - i 4 k_c^3 \left(A_{11}^* \cdot \partial_x A_{22}\right)\right] \\
& + i 2 k_c \left(\Psi_{22} \cdot \partial_x^3 \Psi_{11}^*\right) - i 2 k_c^3\left(\Psi_{22} \cdot \partial_x \Psi_{11}^* \right) - i k_c \left(\partial_x \Psi_{20} \cdot \partial_x^2 \Psi_{11}\right) + i k_c \left(\partial_x \Psi_{22} \cdot \partial_x^2 \Psi_{11}^*\right) \\
& - i k_c \left(\partial_x \Psi_{20}^* \cdot \partial_x^2 \Psi_{11}\right) + i k_c^3 \left(\partial_x \Psi_{20} \cdot \Psi_{11}\right) + i k_c^3 \left(\partial_x \Psi_{20}^* \cdot \Psi_{11}\right) - i k_c^3 \left(\partial_x \Psi_{22} \cdot \Psi_{11}^*\right) \\
& + \Co \left[ - i 2 k_c \left(A_{22} \cdot \partial_x^3 A_{11}^*\right) + i 2 k_c^3 \left(A_{22} \cdot \partial_x A_{11}^*\right) + i k_c \left(\partial_x A_{20} \cdot \partial_x^2 A_{11}\right) - i k_c \left(\partial_x A_{22} \cdot \partial_x^2 A_{11}^*\right) \right] \\
& + \Co \left[ i k_c \left(\partial_x A_{20}^* \cdot \partial_x^2 A_{11}\right) - i k_c^3 \left(\partial_x A_{20} \cdot A_{11}\right) - i k_c^3 \left(\partial_x A_{20}^* \cdot A_{11}\right) + i k_c^3 \left(\partial_x A_{22} \cdot A_{11}^*\right)\right]
\end{split}
\eeq

\beq
\begin{split}
N_{31}^{(u)} = \, & i k_c \left(\Psi_{11} \cdot \partial_x u_{20}\right) + i k_c \left(\Psi_{11} \cdot \partial_x u_{20}^*\right) - i k_c \left(\Psi_{11}^* \cdot \partial_x u_{22}\right) - i 2 k_c \left(\partial_x \Psi_{11}^* \cdot u_{22}\right) \\
& - i k_c \left(u_{11} \cdot \partial_x \Psi_{20} \right) - i k_c \left(u_{11} \cdot \partial_x \Psi_{20}^*\right) + i k_c \left(u_{11}^* \cdot \partial_x \Psi_{22} \right) + i 2 k_c \left(\partial_x u_{11}^* \cdot \Psi_{22}\right) \\
& +\Co \left[- i k_c \left(A_{11} \cdot \partial_x B_{20}\right) - i k_c \left(A_{11} \cdot \partial_x B_{20}^*\right) + i k_c \left(A_{11}^* \cdot \partial_x B_{22}\right) + i 2 k_c \left(\partial_x A_{11}^* \cdot B_{22}\right)\right] \\
& + \Co \left[ i k_c \left(B_{11} \cdot \partial_x A_{20}\right) + i k_c \left(B_{11} \cdot \partial_x A_{20}^*\right) - i k_c \left(B_{11}^* \cdot \partial_x A_{20}\right) - i 2 k_c \left(\partial_x B_{11}^* \cdot A_{22}\right)\right]
\end{split}
\eeq

\beq
\begin{split}
N_{31}^{(A)} =\, & -i k_c \left(A_{11}\cdot\partial_x \Psi_{20}\right) - i k_c \left(A_{11} \cdot \partial_x \Psi_{20}^*\right)
 + i k_c \left(A_{11}^* \cdot \partial_x \Psi_{22}\right) + i 2 k_c \left(\partial_x A_{11}^* \cdot \Psi_{22}\right) \\
 & + i k_c \left(\Psi_{11} \cdot \partial_x A_{20}\right) + i k_c \left(\Psi_{11} \cdot \partial_x A_{20}^*\right) - i k_c \left(\Psi_{11}^* \cdot \partial_x A_{22} \right) - i 2 k_c \left(\partial_x \Psi_{11}^* \cdot A_{22}\right)
\end{split}
\eeq

\beq
\begin{split}
N_{31}^{(B)} = \, & i k_c \left(\Psi_{11} \cdot \partial_x B_{20}\right) + i k_c \left(\Psi_{11} \cdot \partial_x B_{20}^*\right) - i k_c \left(\Psi_{11}^* \cdot \partial_x B_{22}\right) - i 2 k_c\left(\partial_x \Psi_{11}^* \cdot B_{22}\right) \\
& - i k_c \left(B_{11} \cdot \partial_x \Psi_{20}\right) - i k_c \left(B_{11} \cdot \partial_x \Psi_{20}^*\right) + i k_c\left(B_{11}^* \cdot \partial_x \Psi_{22}\right) + i 2 k_c \left(\partial_x B_{11}^* \cdot \Psi_{22}\right) \\
& - i k_c \left(A_{11} \cdot \partial_x u_{20}\right) - i k_c \left(A_{11} \cdot \partial_x u_{20}^*\right) + i k_c \left(A_{11}^* \cdot \partial_x u_{22}\right) + i 2 k_c \left(\partial_x A_{11}^* \cdot u_{22}\right) \\
& i k_c \left(u_{11} \cdot \partial_x A_{20} \right) + i k_c \left(u_{11} \cdot \partial_x A_{20}^*\right) - i k_c \left(u_{11}^* \cdot \partial_x A_{22}\right) - i 2 k_c \left(\partial_x u_{11}^* \cdot A_{22}\right)
\end{split}
\eeq

\section{Linear dispersion relation}\label{app:dispersion}

The linear dispersion relation, which determines the variable scalings in the multiple scales analysis. This relation is found by perturbing the linear system (Equation \ref{eq:ordere}) with a small perturbation of the form $e^{\sigma t + i k_x x + i k_z z}$. Note that the spatial eigenvalues appear as $k_z^2$ and $k_x^2$ at lowest order.

\beq
\begin{split}
& \frac{B_{0}^{4} k_{x}^{2} k_{z}^{4}}{16 \pi^{2}} + \frac{B_{0}^{4} k_{z}^{6}}{16 \pi^{2}} - \frac{B_{0}^{2} \Omega_{0} k_{z}^{4} q}{2 \pi} - 2 \Omega_{0} k_{z}^{2} q \sigma^{2} - \frac{4 \sigma}{\reym} \Omega_{0} k_{x}^{2} k_{z}^{2} q - \frac{4 \sigma}{\reym} \Omega_{0} k_{z}^{4} q - \frac{2 \Omega_{0}}{\reym^{2}} k_{x}^{4} k_{z}^{2} q - \frac{4 \Omega_{0}}{\reym^{2}} k_{x}^{2} k_{z}^{4} q - \frac{2 \Omega_{0}}{\reym^{2}} k_{z}^{6} q \\
& - k_{x}^{2} \sigma^{4} - k_{z}^{2} \sigma^{4} + 4 k_{z}^{2} \sigma^{2} - \frac{2 \sigma^{3}}{\reym} k_{x}^{4} - \frac{4 \sigma^{3}}{\reym} k_{x}^{2} k_{z}^{2} + \frac{8 \sigma}{\reym} k_{x}^{2} k_{z}^{2} - \frac{2 \sigma^{3}}{\reym} k_{z}^{4} + \frac{8 \sigma}{\reym} k_{z}^{4} - \frac{k_{x}^{6} \sigma^{2}}{\reym^{2}} - \frac{3 \sigma^{2}}{\reym^{2}} k_{x}^{4} k_{z}^{2} + \frac{4 k_{x}^{4}}{\reym^{2}} k_{z}^{2} \\
& - \frac{3 \sigma^{2}}{\reym^{2}} k_{x}^{2} k_{z}^{4} + \frac{8 k_{x}^{2}}{\reym^{2}} k_{z}^{4} - \frac{k_{z}^{6} \sigma^{2}}{\reym^{2}} + \frac{4 k_{z}^{6}}{\reym^{2}} - \frac{2 \sigma^{3}}{\reye} k_{x}^{4} - \frac{4 \sigma^{3}}{\reye} k_{x}^{2} k_{z}^{2} - \frac{2 \sigma^{3}}{\reye} k_{z}^{4} - \frac{4 k_{x}^{6} \sigma^{2}}{\reye \reym} -  \frac{12 k_{x}^{4} k_{z}^{2} \sigma^{2}}{\reye \reym} - \frac{12 k_{x}^{2} k_{z}^{4} \sigma^{2}}{\reye \reym} \\
& - \frac{4 k_{z}^{6} \sigma^{2}}{\reye \reym} - \frac{2 k_{x}^{8} \sigma}{\reye \reym^{2}} - \frac{8 k_{x}^{6} k_{z}^{2} \sigma}{\reye \reym^{2}} - \frac{12 k_{x}^{4} k_{z}^{4} \sigma}{\reye \reym^{2}} - \frac{8 k_{x}^{2} k_{z}^{6} \sigma}{\reye \reym^{2}} - \frac{2 k_{z}^{8} \sigma}{\reye \reym^{2}} - \frac{k_{x}^{6} \sigma^{2}}{\reye^{2}} - \frac{3 \sigma^{2}}{\reye^{2}} k_{x}^{4} k_{z}^{2} - \frac{3 \sigma^{2}}{\reye^{2}} k_{x}^{2} k_{z}^{4} - \frac{k_{z}^{6} \sigma^{2}}{\reye^{2}} \\
& - \frac{2 k_{x}^{8} \sigma}{\reye^{2} \reym} - \frac{8 k_{x}^{6} k_{z}^{2} \sigma}{\reye^{2} \reym} - \frac{12 k_{x}^{4} k_{z}^{4} \sigma}{\reye^{2} \reym} - \frac{8 k_{x}^{2} k_{z}^{6} \sigma}{\reye^{2} \reym} - \frac{2 k_{z}^{8} \sigma}{\reye^{2} \reym} - \frac{k_{x}^{10}}{\reye^{2} \reym^{2}} - \frac{5 k_{x}^{8} k_{z}^{2}}{\reye^{2} \reym^{2}} - \frac{10 k_{x}^{6} k_{z}^{4}}{\reye^{2} \reym^{2}} - \frac{10 k_{x}^{4} k_{z}^{6}}{\reye^{2} \reym^{2}} \\
& - \frac{5 k_{x}^{2} k_{z}^{8}}{\reye^{2} \reym^{2}} - \frac{k_{z}^{10}}{\reye^{2} \reym^{2}} = 0
\end{split}
\eeq

\end{document}